%% file: paper.tex
\documentclass{svproc}
\usepackage{url}
\usepackage{graphicx}
\usepackage{epsfig}
\usepackage{xspace}
\usepackage{wrapfig}

\newcommand\blfootnote[1]{%
  \begingroup
  \renewcommand\thefootnote{}\footnote{#1}%
  \addtocounter{footnote}{-1}%
  \endgroup
}\input{commands.tex}

\begin{document}
\mainmatter              
\title{Studying the effect of the hadronic phase in nuclear collisions with PYTHIA and UrQMD}
\titlerunning{Studying the hadronic phase in PYTHIA+UrQMD}  
%
\author{A.S.~Vieira\inst{1}, C.~Bierlich\inst{2}
D.D.~Chinellato\inst{1}, 
J.~Takahashi\inst{1}}
\authorrunning{A.S.~Vieira et al} 
%
\tocauthor{David Dobrigkeit Chinellato}
\institute{Universidade Estadual de Campinas, S\~{a}o Paulo, Brazil
\and
Niels Bohr Institute, University of Copenhagen, Denmark}

\maketitle 

\begin{abstract}
In this work, we couple Pb-Pb events simulated with the PYTHIA Angantyr event generator at $\sqrt{s_{\rm{NN}}}=2.76$ and $5.02$~TeV with the hadronic cascade simulator UrQMD to study the effect of the hadronic phase on observables such as charged-particle multiplicity densities, transverse momentum spectra and identified particle ratios, giving special emphasis to short-lived resonances. 
\end{abstract}
\blfootnote{CB was supported by Swedish Research Council, contract number 2017-003. AV, DDC and JT were supported by FAPESP grant 2017/05685-2.}
The extreme conditions reached in ultra-relativistic heavy-ion collisions at the LHC are expected to produce a state of matter in which quarks and gluons are deconfined, the quark-gluon plasma (QGP). As a consequence, several features, such as elliptic flow and chemically equilibrated particle production, are expected and observed in these collision systems. However, it has to be noted that, once hadronization takes place, inelastic and elastic interactions may still take place. A proper distentanglement of the effects of this final hadronic phase and any features emerging from previous stages of the system evolution is fundamental to the understanding of heavy-ion collisions. 

\section{The used models}
\label{sec:angantyr}
To characterize the hadronic phase, we 
couple the PYTHIA Angantyr model \cite{Bierlich:2016smv,Bierlich:2018xfw} to the hadronic cascade
simulator UrQMD \cite{Bleicher:1999xi}. The Angantyr model 
is based on the wounded nucleon model \cite{Bialas:1976ed}, 
and as such on a Glauber model. In Angantyr, 
several additions to the original model has been made. 
The full nucleon--nucleon scattering amplitude is
parametrized, allowing for a distinction between nucleons which are elastically
scattered, diffractively excited or participating with colour exchange from
both nucleons, denoted absorptively scatterred. All parameters
in the parametrization of the scattering amplitude are fitted 
to the total and semi-inclusive cross sections in pp collisions. 
The parameters of the PYTHIA models for multiparton interactions (MPIs) \cite{Sjostrand:1987su}, 
hadronization, parton showers as well as parton density
functions, are all fixed in $e^+e^-$, $e$p and pp collisions.
In a pp collision, MPIs are selected by the $2\rightarrow 2$ perturbative
parton--parton scattering cross section. Since this cross section has a $1/p^4_\perp$
divergence, it is regulated by one of the aforementioned parameters of the MPI model.
In a heavy ion collision, a single projectile nucleon can interact absorptively
with several target nucleons. In such cases, shadowing effects must be taken into account.
In Angantyr this is done by ordering absorptively wounded nucleons into two categories, using
the following strategy. All \textit{possible} nucleon--nucleon interactions are ordered in 
increasing nucleon--nucleon impact parameter. Going from smallest to largest impact parameter,
the nucleon pairs are labelled \textit{primary} if neither of the nucleons have participated before,
and \textit{secondary} if one of the nucleons have previously participated in an interaction. The primary
interactions are modelled as a normal inelastic non-diffractive collision. A secondary interaction is treated 
as only a single wounded nucleon, with inspiration from the Fritiof model \cite{Andersson:1986gw}, where a wounded nucleon
contributes to the final state as a string with a mass distribution $\propto dM^2/M^2$, similar as 
the modelling of diffractive excitation. Instead of using a single string, as the original Fritiof
program did, Angantyr allows for MPIs in secondary collisions, by treating the system as a collision
between a nucleon and an Ingelman--Schlein Pomeron \cite{Ingelman:1984ns}, followed by a (perturbative) parton shower.
The nucleon--Pomeron scatterings are adjusted to look as normal inelastic non--diffractive nucleon--nucleon
scatterings in the direction of the wounded nucleon.
Furthermore, secondary collisions are rejected in cases where they would break overall energy--momentum conservation. This has the effect that not all possible nucleon--nucleon interactions from the Glauber calculation,
are in fact realized in the generation of a final state.

\begin{figure}
\begin{center}
\includegraphics[width=0.49\textwidth]{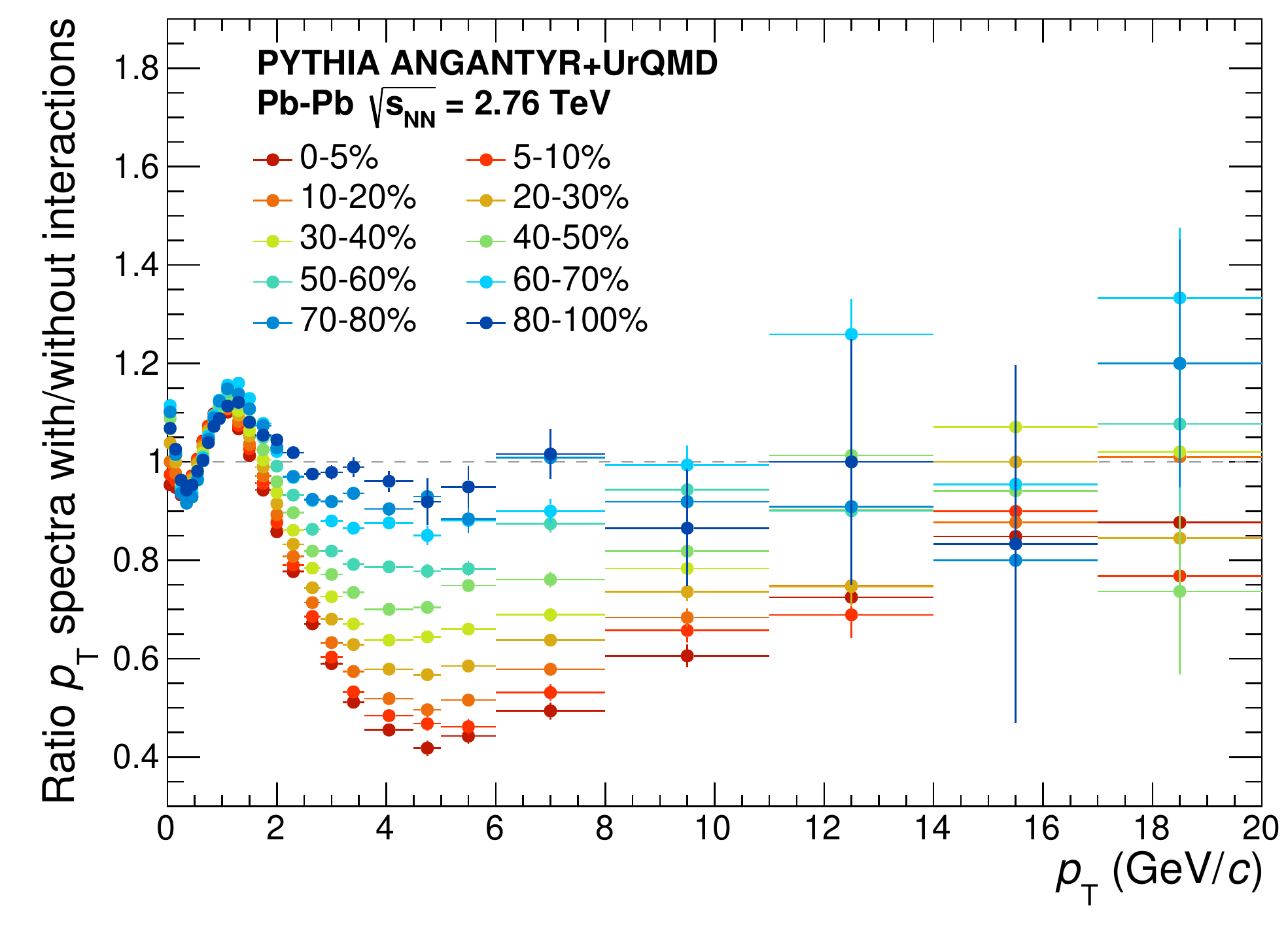}
\includegraphics[width=0.49\textwidth]{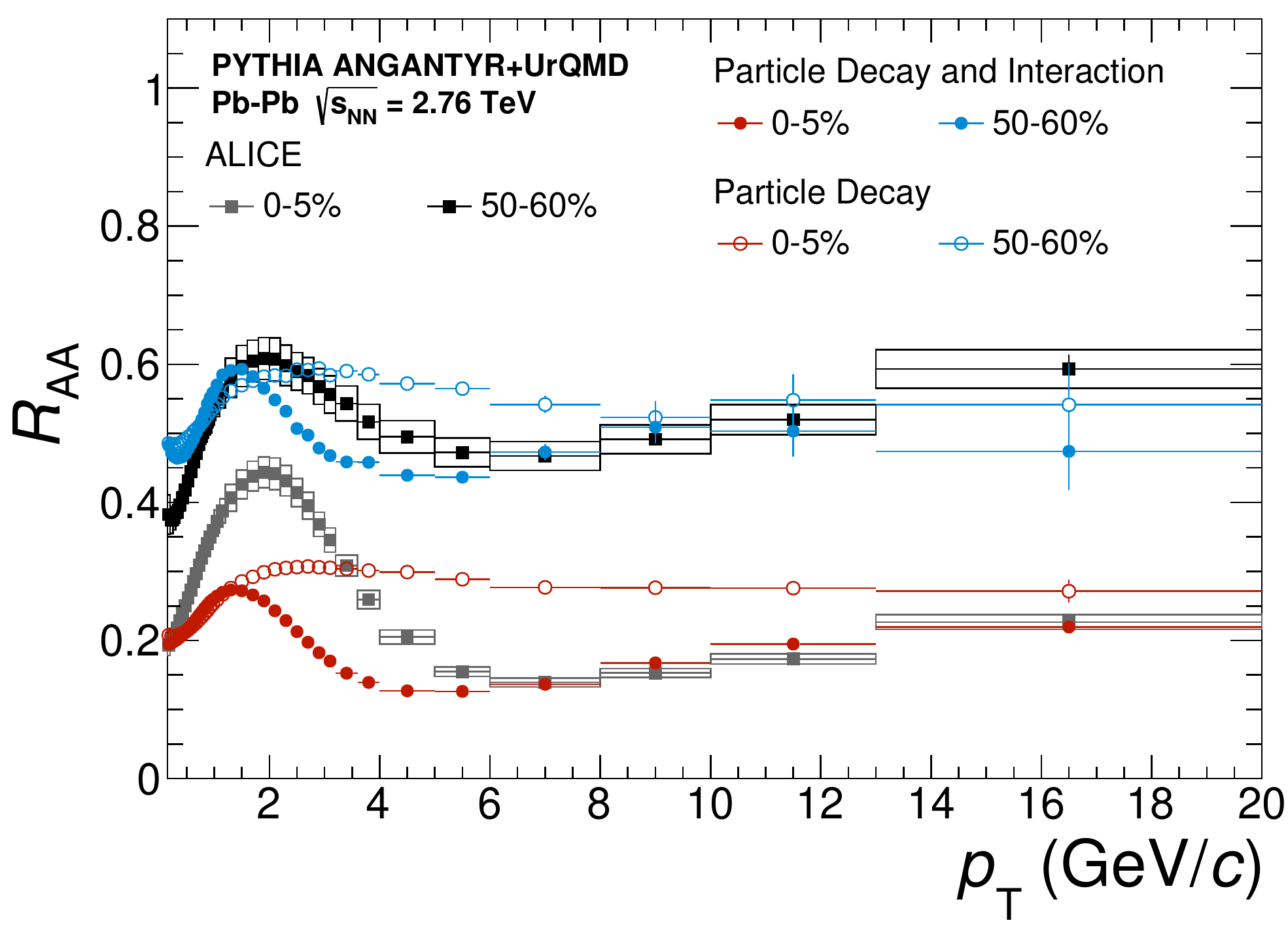}
\label{fig:modification}
\caption{(left) Ratio of \pt spectra with hadronic interactions and
without interactions in various centrality classes and (right) $R_{\rm{AA}}$ in two selected centrality classes in Pb-Pb at 2.76 TeV. Data on the right plot are from \cite{Acharya:2018qsh}.}
\end{center}
\end{figure}
As such, the Angantyr framework produces hadronic final states of heavy ion collisions (p$A$ and $AA$), without any
free parameters, as all are fixed by data from smaller collision systems. It should be noted that this treatment 
generates a heavy ion final state with no effects from a potential QGP, thus providing a baseline for tests of various
models for collectivity. The recent models for ``Rope hadronization'' \cite{Bierlich:2014xba} and ``String shoving'' \cite{Bierlich:2017vhg} implemented in
Pythia, are not included in any results of these proceedings.
The final state hadrons are assigned spatial vertices calculated in
the string model \cite{Ferreres-Sole:2018vgo}. 
These spatial vertices are then passed on to UrQMD, 
leading to a complete model of a heavy-ion collision in which 
no equilibrium is assumed. 

\section{Results}
In order to test if this simulation chain describes heavy-ion
collision data adequately, we compare predictions of the 
basic observables \dndeta and \meanpt to ALICE measurements \cite{Aamodt:2010cz}. While the \dndeta is within 10\% of the measured values, the \meanpt exhibits stronger deviations due to the absence of 
radial flow. The corresponding plots have been omitted for brevity. 

\begin{wrapfigure}{r}{0.54\textwidth}
\includegraphics[width=0.54\textwidth]{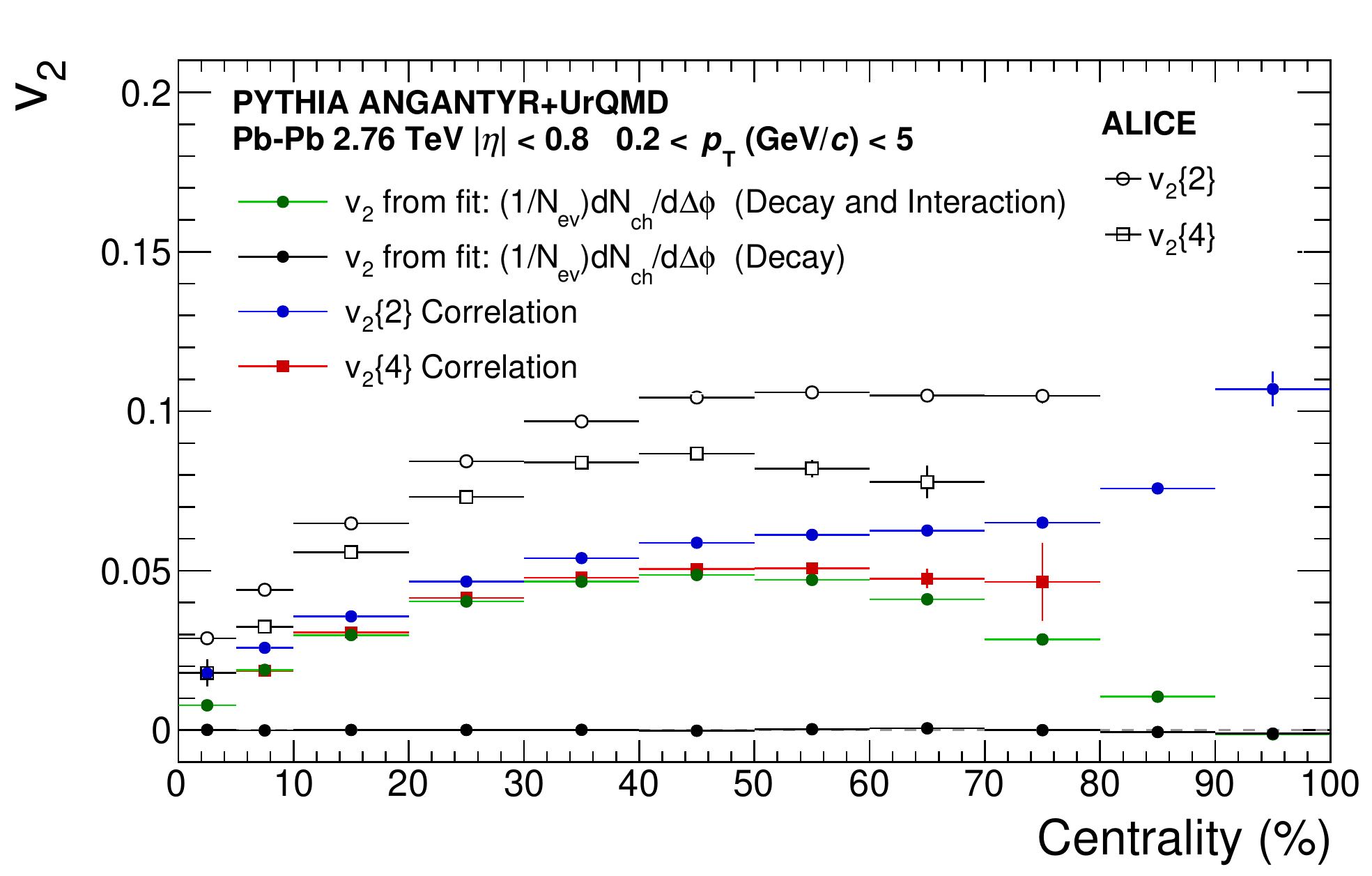}
\label{fig:ellipticflow}
	\caption{The flow coefficients $v_2\{2\}$ and $v_2\{4\}$ from Pythia8/Angantyr + UrQMD (blue and red), as well as $v_2$ extracted from fit of the $dN_{ch}/d\Delta\phi$ distribution, all as function of centrality. Compared to data from ALICE \cite{Aamodt:2010pa}.}
\end{wrapfigure}

To further characterize the effect of the hadronic phase, 
we study how the \pt-differential spectra of charged 
particles are altered by hadronic scattering in UrQMD by 
calculating the ratio of the \pt spectra obtained when 
scattering is allowed to the spectra obtained when only decays 
occur in the hadronic phase. As can be seen in Fig.~\ref{fig:modification}, hadronic interactions 
are seen to reduce high-\pt yields rather significantly, with 
the effect being most pronounced around 5~GeV/$c$. In addition, 
a minor radial-flow-like effect can be seen at low-\pt, where
very low momentum particles are pushed to a momentum of about 1.5~GeV/$c$. 

The observation that high-\pt is suppressed in this way prompts
the calculation of the nuclear modification factor $R_{\rm{AA}}$, 
shown in Fig.~\ref{fig:modification} for two selected centralities and for PYTHIA+UrQMD calculations with and without hadronic interactions. For this calculation, the same
framework was used to determine particle spectra in pp collisions at
$\sqrt{s}$ = 2.76~TeV and the number of binary collisions is taken 
to be the one calculated by the ALICE experiment \cite{ALICE-PUBLIC-2018-011}. 
While for low \pt the $R_{\rm{AA}}$ from simulations fails
at describing the measurement, the model matches data
above a \pt of 6~GeV/$c$. Quantitatively, two contributions lead
to this correct description of the $R_{\rm{AA}}$:\\
\textbf{a)} Angantyr does not follow normal $N_{coll}$ scaling. As explained in section \ref{sec:angantyr}, the number of binary collisions does not enter as a scaling parameter, as shadowing effects are included by treatment of secondary wounded nucleons.\\
\textbf{b)} UrQMD further modifies the $R_{\rm{AA}}$, through rescattering effects, to precisely reproduce the minimum 
at around 5-6~GeV/$c$.\\ 
Furthermore, the model 
is also seen to exhibit a progressively smaller high-\pt 
suppression for increasing \pt. In the framework of
the hadron vertex model, this is understood to be a consequence
of high-momentum particles originating from boosted 
jet-like structures that hadronize progressively further away
from the bulk of particles created in a heavy-ion collision.
Another observable commonly used to characterize heavy-ion 
collisions is the anisotropic flow, quantified via the
flow coefficients $v_{n}$. The leading elliptic flow term, $v_{2}$, 
is shown in Fig.~\ref{fig:ellipticflow} as a function 
of centrality in Pb-Pb collisions at 2.76~TeV simulated with the
PYTHIA+UrQMD chain. The $v_{2}$ has been calculated in two distinct ways: 
first, with a fit to the azimuthal particle distribution 
with respect to the simulated event plane $dN_{ch}/d\Delta\phi$, 
and then also using 2- and 4-particle cumulants following
ref.~\cite{cumulantsBilandzic2011}.
If no hadronic interactions take
place, particle production is seen to exhibit no correlation with
the event plane as expected. However, if hadronic scattering is 
enabled, a substantial amount of elliptic flow builds up in the 
hadronic phase, reaching around 50-60\% of the values measured
by ALICE \cite{Aamodt:2010pa}, which is consistent with observations
from earlier studies using UrQMD \cite{Lu_2006}. 
The elliptic flow estimated via 4-particle
cumulants $v_{2}\{4\}$ matches the one calculated via a fit to 
the $dN_{ch}/d\Delta\phi$ distribution in 0-50\% collisions, 
indicating that it is well correlated with initial hadronic geometry, while for more peripheral events this correlation 
with the simulated event plane becomes weaker, indicating that
the non-zero $v_{2}\{2\}$ and $v_{2}\{4\}$ in peripheral events
is due to non-flow contributions such as dijet structures. 
Therefore, even in the absence of initial hadronic
flow, hadronic interactions may build up a substantial amount of 
elliptic flow. Further studies are needed to determine the 
exact level of flow that would have to be present at hadronization 
time so as to be able to reproduce measurements from ALICE. 

%
\bibliographystyle{mybib}   
\bibliography{bibliography}

\end{document}

%% file: commands.tex
\newcommand{\pt}           {\ensuremath{p_{\rm T}}\xspace}
\newcommand{\meanpt}       {$\langle p_{\mathrm{T}}\rangle$\xspace}

\newcommand{\dndeta}       {\ensuremath{\mathrm{d}N_\mathrm{ch}/\mathrm{d}\eta}\xspace}

\newcommand{\nineH}        {$\sqrt{s}~=~0.9$~Te\kern-.1emV\xspace}
\newcommand{\seven}        {$\sqrt{s}~=~7$~Te\kern-.1emV\xspace}
\newcommand{\twoH}         {$\sqrt{s}~=~0.2$~Te\kern-.1emV\xspace}
\newcommand{\twosevensix}  {$\sqrt{s}~=~2.76$~Te\kern-.1emV\xspace}
\newcommand{\five}         {$\sqrt{s}~=~5.02$~Te\kern-.1emV\xspace}
\newcommand{\twosevensixnn}{$\sqrt{s_{\mathrm{NN}}}~=~2.76$~Te\kern-.1emV\xspace}
\newcommand{\fivenn}       {$\sqrt{s_{\mathrm{NN}}}~=~5.02$~Te\kern-.1emV\xspace}

\newcommand{\GeVc}         {Ge\kern-.1emV/$c$\xspace}
\newcommand{\MeVc}         {Me\kern-.1emV/$c$\xspace}
\newcommand{\TeV}          {Te\kern-.1emV\xspace}
\newcommand{\GeV}          {Ge\kern-.1emV\xspace}
\newcommand{\MeV}          {Me\kern-.1emV\xspace}
\newcommand{\GeVmass}      {Ge\kern-.2emV/$c^2$\xspace}
\newcommand{\MeVmass}      {Me\kern-.2emV/$c^2$\xspace}

\usepackage{xcolor}

\usepackage[colorlinks]{hyperref}
\hypersetup{
  colorlinks=true, 
  linktoc=all,     
  linkcolor=black,  
  citecolor= blue, 
  urlcolor  = blue 
}